# Multi-phase IRC Botnet and Botnet Behavior Detection Model

Aymen Hasan Rashid Al Awadi
Information Technology Research Development Center, University of Kufa, Najaf, Iraq
School of Computer Sciences Universiti Sains Malaysia 11800 USM, Penang, Malaysia

Bahari Belaton
School of Computer Sciences Universiti Sains Malaysia 11800 USM, Penang, Malaysia

## ABSTRACT

Botnets are considered one of the most dangerous and serious security threats facing the networks and the Internet. Comparing with the other security threats, botnet members have the ability to be directed and controlled via C&C messages from the botmaster over common protocols such as IRC and HTTP, or even over covert and unknown applications. As for IRC botnets, general security instances like firewalls and IDSes do not provide by themselves a viable solution to prevent them completely. These devices could not differentiate well between the legitimate and malicious traffic of the IRC protocol. So, this paper is proposing an IDS-based and multi-phase IRC botnet and botnet behavior detection model based on C&C responses messages and malicious behaviors of the IRC bots inside the network environment. The proposed model has been evaluated on five network traffic traces from two different network environments (Virtual network and DARPA 2000 Windows NT Attack Data Set). The results show that the proposed model could detect all the infected IRC botnet member(s), state their current status of attack, filter their malicious IRC messages, pass the other normal IRC messages and detect the botnet behavior regardless of the botnet communication protocol with very low false positive rate. The proposed model has been compared with some of the existing and well-known approaches, including BotHunter, BotSniffer and Rishi regarding botnet characteristics taken in each approach. The comparison showed that the proposed model has made a progress on the comparative models by not to rely on a certain time window or specific bot signatures.

## 1. INTRODUCTION
Over the last decade, the botnets become the most dangerous threats that can threaten the existing services and resources over the Internet. Many malicious activities could be performed for exploiting the victims (bots) to attack another victim. The bot itself is malicious software like the common computer viruses and worms. The bots could be distinguished from the other malicious software by implementing Command and Control (C&C) directions, which are set of messages that can be used by the botmaster to direct and control the connected bots through certain connection channel (IRC or HTTP). In this way, the attacker will be able to evade himself by exploiting the bots to attack another victim. However, the bots could be found in any environment like in home, schools, banks and any of governmental institutes making use of system vulnerabilities and software bugs to separate and execute a lot of malicious activities. Recently, bots can be the major one of the major sources for distributing or performing many kinds of scanning related attacks (Distributed Denial-of-Service DoS) [1], spamming [2], click fraud [3], identity fraud, sniffing traffic and key logging [4] etc. The nature of the bots activities is to respond to the botmaster's control command simultaneously. This responding will enable the botmaster to get the full benefit from the infected hosts to attack another target like in DDoS [5]. From what stated earlier, the botnet can be defined as a group of connected agents (bots) controlled by certain botmaster and can perform similar communication pattern and malicious attacks toward certain victim(s) [6]. Mainly, botnets could be characterized into two types; the first one is the centralized architecture whereby all the bots will connect and controlled by certain botmaster using IRC (Internet Relay Chat) [7] and HTTP. In IRC-based architecture, the botmaster will interact with the bots directly and in real time manner using (IRC PRIVMSG), while in HTTP-based the bot will periodically connect to the C&C server to obtain the command in a centralized way [7]. The second architecture of botnet architecture is P2P (Point to Point) architecture. This architecture does not have a central C&C server, and all the bots will be connected to each other to get controlled. Because of the property of not having any centralized C&C server, P2P botnet does not suffer from a single point of failure [8]. However, the central architecture is proven more flexible to the attackers. Since it provides instant interaction with dozens of zombies (bots) efficiently, therefore the advantage of the botmaster will be maximized by exploiting all of those bots [7,9].

## 2. RELATED WORKS
Defending against botnet activities is focusing on detection and monitoring, prevention and mitigation botnet activities that outgoing from the internal environment network. According toZeidanloo and Manaf in [10], botnet detection and monitoring can be classified into two main categories, which are honeynet analysis based and IDS-based. In IDS-based, the detection could be classified into anomaly and signature based. The anomaly-based botnet detection can also subdivided into anomaly host-based and anomaly network-based.

Honeynet is a group of instances that have real systems, applications and services with no production and authorization activities [11]. So, any outgoing or incoming connection (C&C of IRC or HTTP) to these systems will be captured and analyzed as a suspicious activity. Many research papers have addressed botnet detection depending on honeynet. Nepenthes, which is a honeypot-based framework has been proposed by Baecher et al. in [12] for collecting self-replication malware that use a variety of protocols including; TFTP, FTP, IRC, HTTP and custom protocols, all in wide-scale environment. The framework virtualized only the vulnerable services in the honeypot for more efficiency. Good deployment strategy has been proposed in the framework, as a precautionary measure to prevent the framework from getting exploited. Botnet behavior study is one of the most important methodologies to understand botnet phenomenon.Zhichun et al.have proposed a honeynet-based framework for botnet detection depending on botnet scanning behavior [13]. Honeynet has been used to detect the scanning activities of the bots. The authors decided to depend on such activity since the scanning process can be considered the initial step to maintain the availability of the bot.

IDS has been employed to detect bots and their activities since it is considered one of the devices that used to accomplish defense in depth strategy along with IPS (Intrusion Prevention System) and firewall. The main purpose of the IDS is to raise an alarm on detecting malicious or anomalous activity within network traffic. Mainly, IDS could be classified according to the method that been used to detect the malicious activity to: signature-based and anomaly-based IDS. Signature-based method can only detect the verified and known signatures, meaning to say that zero-day botnet activity will not be detected. Goebel and Holzhave proposed Rishiin [14], which is a passive monitoring botnet detection approach based on IRC signatures. This approach exploits n-gram analysis and scoring system to detect the suspicious IRC nickname's patterns, IRC server and uncommon server ports in the monitored network, for giving an evidence for botnet existence. In the large-scale network, identifying application communications could be a good way to identify normal behavior (made by human) from the anomalous behavior of bots [15]. This approach has been exploited byLiuet al. in[15], where the network traffic of known applications will be identified depending on signature-based approach of each application. N-gram features also extracted and clustered to verify the anomalous behavior of the identified application. Anomaly-based approach depends on monitoring certain behaviors to detect the anomalous activities. Many properties can be monitored to give an evidence of botnet existing like high traffic on unusual ports, unusual high traffic, high consuming bandwidth and high latency in the network [6]. Binkley and Singh proposed an algorithm based on anomaly strategy to detect IRC botnet clients and server inside the monitored network [16]. The proposed algorithm combines two tuples, IRC botnet detection along with TCP SYN scanning detection heuristic. IRC tuple produced two tuples, one for distinguishing IRC based on the IP channel name, and the other sub-tuple is responsible for providing TCP work weight on individual IRC channels.

Host-based is kind of anomaly strategy that is depending on monitoring botnet behaviors via the host. Stinson and Mitchell have proposed host-based analysis approach to detect the botnets command behaviors called BotSwat [17]. Mainly, this approach depends on distinguishing the remote invocations calls that issued by the botmaster to the bots from the local invocations calls that could be issued legitimately. In network-based strategy, the network flow will be considered in the process of botnet and botnet behavior detection. In this strategy, there are two ways which are network-based active monitoring and network-based passive monitoring. Gu et al. proposed BotProbe[18], which is an efficient approach based on active monitoring, where the IRC C&C interactions of bots distinguished from common IRC messages issued by human based on C&C signatures. The authors proposed a hypothesis testing framework based on cause and effect approach for probing behaviors that can happen when sending additional packets to the suspicious hosts and observing its response for several times. BotSniffer byGu et al. depended on the network passive monitoring approach [7]. This method is an anomaly-based approach focusing on the Spatial-temporal and similarity correlation mechanism for the C&C messages (IRC-based and HTTP-based) botnets and other botnet activities within a certain time window.

In this paper, the proposed work touches certain research areas related to botnet detection techniques and network monitoring approaches to obtain the required features for IRC botnet and botnet behavior detection. So, itcould be classified as an IDS-based botnet detection model that is depending on anomaly passive network monitoring and IRC responses messages behavior for detecting IRC bot(s). As for botnet behavior, the proposed model is also IDS-based approach but with network active monitoring and botnet behavioral based for the detected attacks that achieve the definition of botnet attacks.

## 3. The Proposed Approach

IDS is a powerful tool that can be used to monitor botnet member's activities by detecting IRC responses messages of the bots through IRC channel and their outgoing attacks. However, the problem with IDS alerts is the poor quality. The non-well optimized sensor could produce a lot of false alerts, especially for the botnet attacks that could make the alert's analysis process inefficient. The input lines of the proposed model will be from the raw events of the IDS alerts. So, a way to filter out and cluster the alerts has been proposed whereby the false alerts could be reduced as possible. In the upcoming sections, the main used methods beside the phases of botnet detection will be addressed.

### 3.1 Alerts Correlation

IDS alerts raised for wide ranges of network malicious activities. There are several alerts' attributes of those activities, including (source IP, destination IP, timestamp, signature name, etc.). These attributes could be similar or different regarding to the situation of the attack. So, studying and analyzing the relationships (correlations) and similarities between these attributes as a group could provide efficient information about certain activity without examining the alerts individually. Dingbang and Peng in [19] classified alerts correlation methods into four methods:
1- Finding the similarities between the alerts' attributes.
2- Matching the alerts according to a predefined attack scenario.
3- Matching alerts' attributes depending on causes and effects (preconditions and consequences) of certain attack's scenarios, where the effects of those attacks will be matched with the causes that could produce such effects.
4- The last method will depend on retrieving information from multiple sources and integrating them together with IDS alerts.
In the proposed model,the used correlation method is spatial-temporal correlation method [7], which is a special case of the first method of correlation methods (alerts similarities).

### 3.2 Spatial-Temporal Correlation and Similarities Method

Since the observed behaviors could be represented through IDS alerts, so an efficient method should be used for analyzing these alerts lead to infer on the existence of botnet. The spatial-temporal correlation method is trying to find the relationship between the spatial alerts' attributes which are (signature name, source IP, source port, destination IP and destination port) and the temporal attribute which is the timestamp value. This method will be exploited to analyze all alerts related to IRC protocol that having similar attributes. Then the alerts will be clustered based the similarity method. The type of that clustering will be used over all the phases of the proposed model, as will be clarified in the upcoming sections.

## 3.3 The Multi-phase IRC Botnet and Botnet Behavior Detection Model Algorithm

In this paper, the botnet behaviors will be modeled based on its phases in the life-cycle. So, the proposed model will consist of two phases for IRC botnet and botnet behavior detection (Phase 1 and Phase 2). Where Phase 1 is representing Phase 2 of botnet life-cycle (communication phase), while Phase 2 is representing phase 3 of botnet life-cycle (attacking phase). Figure 1 is depicted the proposed model's components and its stages.

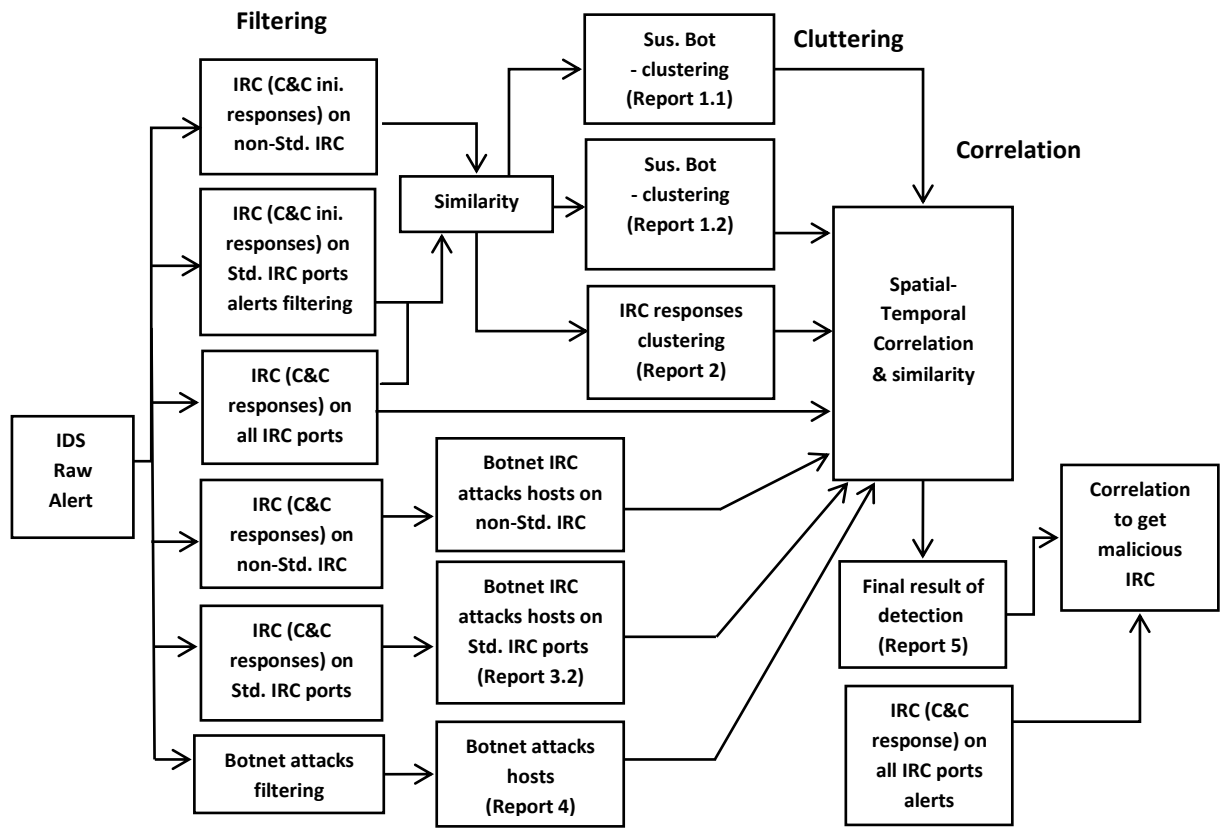

**Fig 1: The proposed model for IRC botnet and botnet behavior detection**

## 4. The Proposed phases and the Process Flows Modes of the Model

There are two phases in the proposed model, the first phase of the model will not represent the first phase of the botnet life-cycle since the used methods in this phase are varying between exploiting the operating systems and services' vulnerabilities to the exploitation of social engineering tactics [20]. However, the first phase will be responsible for collecting all initial outgoing IRC C&C responses messages of the bot in case of standard or non-standard IRC ports. The second phase will deal with the botnet behaviors, where these behaviors will be divided into IRC replies and other detected botnet attacks like DDoS. The proposed phases also will be divided into parts, where every part will represent certain activity in the proposed process flows (coherent or non-coherent). As for the process flow of the model, there are two process flows in the proposed model. The first one is the coherent mode. In this phase the focusing will be on the IRC responses messages which are related to the verified initial IRC bot activities on any port. The implementation of this flow will show that the detected bot was monitored from its initial activity. Non-coherent mode is the second flow, where the IRC responses messages which are not related to the initial IRC activity will be treated. The upcoming subsections will address in details the implementation of the proposed phases and the implicit stages of filtering, clustering and correlation beside the expected results within each one.

### 4.1 Phase 1

The first phase of the model is responsible for collecting the initial and suspicious IRC C&C bot connections in coherent mode process flow. The suspicious IRC bot connections on the non-standard IRC ports will be in Part A and the IRC bot connections on the standard defined IRC ports (6661-6668) will be in Part B. In the other side, the IRC bot responses to the botmaster on any IRC ports will be in Part 2 of this phase. To achieve this phase and all of its parts, three IRC signature rules have been exploited from default Snort signature rules in Snort version (2.8.5.2). In Part 1 (both Part A and B), two rules have been used, one from [21] for monitoring IRC

NICK changes at the initial IRC bot connectivity and for tracking the destination port of the botmaster IRC C&C server. The second one is for detecting the IRC responses messages (PRIVMSG) on the tracked IRC port. The Part 2 of this phase is achieved by implementing a simple rule to detect the ordinary IRC PRIVMSG on any port. The standard range of IRC ports has been assumed based on Bleeding Snort rule (bleeding-attack_response.rules), which is assuming that every IRC PRIVMSG coming from every port except the range (6661-6668) is an IRC attack response on non-standardIRC port [22]. The defined range of standard IRC ports is flexible and can be changed to accommodate the desired network policy rules regarding to the IRC service policy. Both of the standard and non-standard IRC ports will be treated in the proposed model, but in a different way. However, every part of this phase will be implemented concurrently with the three stages (filtering, similarity and clustering) as follows:

### 4.1.1 Phase 1 - Part 1
This part is achieving by employing a set of Snort signature rules proposed by Bianco[21]. However, these rules are suffering from false positive results on IRC botnet detection. Since they cannot distinguish between legitimate and illegitimate IRC connections. The problem of false positive IRC botnet will be handled on the filtering stage of this part. Bianco has proposed 3 rules to detect the initial IRC behavior for the bots. The first rule (rule 1) is responsible for monitoring and tagging the IRC bot connection request and then preserves the destination port of the IRC C&C server for connection monitoring purpose. Rule 2 is responsible for monitoring any response that comes from the monitored IRC C&C server on the monitored IRC destination port. Rule 3 (the added rule) is responsible for detecting the outgoing IRC PRIVMSG in any port number, and this response will be on the same IRC destination port that has been tagged already by rule 1. In the proposed model, only rule 1 and rule 3 will be used to track the initial IRC bot behaviors stated above. The implementation of the first part of Phase 1 will be divided into two parts. Both of them will be implemented by implementing the three stages respectively (filtering, similarity and clustering). Firstly, Part A will be implemented by filtering the raw alerts and takes only the alerts that related to the initial IRC responses messages that can be represented by the rule 3 on the non-standard IRC ports. As for Part B, only alerts that belong to rule 3 on the standard IRC port (6661-6668) will be collected. In the second stage, similarity stage, where similar alerts that belong to rule 3 and having similar alerts attributes like (timestamp and source port) that appear in both Part 1 (Part A and Part B), and Part 2 will be kept in Part 1(Part A or Part B) and omitted from Part 2. Since rule 3 has been added to track IRC message responses in coherent mode and the other rule in Part 2 also represent the IRC message responses (PRIVMSG) but for the non-coherent mode. The clustering stage is different in each part, where in Part A alerts' attributes like (timestamp, source IP, source port, destination IP and destination port) will be clustered for every alert on the non-standard IRC port. On the other hand, for Part B, the same attributes will be taken but for more than one IRC bot acting at the same timestamp on the standard IRC port. The clustered alerts in Part A and Part B will extract the bots that are responsible for the initial IRC behaviors. These bots will be saved in Report 1.1 and get status message 1 "Coherent mode: IRC bot has illegitimate IRC connection on non-standard IRC port". As for Part B, its results will be saved in Report 1.2 and get status message 2 "Coherent mode: IRC bot has illegitimate IRC connection on standard IRC port".

### 4.1.2 Phase 1 - Part 2
The second part of Phase 1 will be responsible for detecting the outgoing C&C responses of the bots to the botmaster on any range of IRC ports (non-coherent mode). Again in this part, the main stages (filtering, similarity and clustering) will be performed to get the bots that have similar IRC response messages to a certain IRC C&C server (botmaster) at the same timestamp. In the filtering stage, all alerts which are representing outgoing IRC responses (PRIVMSG) will be collected. The collected alerts will be filtered in a way to ensure that the responses are happening as C&C responses messages to a botmaster C&C. This kind of filtering will lookfor similar alerts having similar alerts attributes like (timestamp, destination IP, destination port) for more than one IRC bot working together at the same timestamp. At this stage, the timestamp value of first outgoing IRC PRIVMSG that appears in the tested network traffic will be preserved as (*time_log*) for filtering purpose in Phase 2. Similarity stage between Part 1 (Part A and Part B) and Part 2 will be implemented to ensure that Part 2 will take only the alerts that belong to the non-coherent mode. The similarity stage will be achieved by matching alerts' attributes such as (timestamp, source port) that appears on both filtering alerts of Part 1(Part A and Part B) and Part 2. Finally, the clustering stage will be performed to extract the IRC bots that are performing IRC responding activities, by clustering (source IP, source port, destination IP and destination Port) for more than one IRC bot. These bots will be saved in Report 2 and get status message 9 "Non-coherent mode: IRC bot has illegitimate IRC connection". In coherent mode, Report 1.1 and Report 1.2 are ready now to be correlated with the second phase of the proposed model. Report 1.1 also could be completely empty, if the Snort signature rules failed for any reason to detect the IRC connection request and responses. In such a case, the proposed model will not consider Part 1 at all and skip it to Part 2. As for the result of IRC botnet detection, it will be in non-coherent mode.

## 4.2 Phase 2
This phase is representing the attacking phase of the botnet's life-cycle, so it will deal with the alerts of the outbound attacking traffic. This part will be divided into two parts (Part 1 and Part 2). Each part will be responsible for a kind of botnet attack. The first part is responsible for IRC attacks (information stealing through the IRC channel of the bot). The second part will be designated for the detected botnet attacks like DDoS attacks. The first part of this phase will be divided into two parts (Part C and Part D), where Part C will be for IRC replies on non-standard range of IRC ports, and Part D will be for IRC replies on the standard range of IRC ports.

### 4.2.1 Phase 2 - Part 1
As stated earlier, the Part 1 will be divided into two parts (Part C and Part D). Part C will collect all outgoing alerts that represent IRC responses messages (C&C responses) on the non-standard IRC ports to the IRC C&C server. In the process of filtering, these alerts will be collected with timestamp longer than the preserved *(time_log)* to ensure that there are continuous C&C responses between the detected IRC bot and its botmaster at the same IRC channel. The filtered alerts will be clustered according to alerts' attributes like (source IP, source port, destination IP and destination port). This part will be used also to detect the activity of the one IRC bot that works on the non-standard IRC port. So, the clustered IRC bot will get status message 7 "IRC bot has illegitimate IRC connection on non-standard IRC port". As for Part D, the filtering procedure will be the same as in Part C, but for the C&C responses on the standard range of IRC ports (6661-6668). This part will not be clustered, since the alerts of this part will be used for correlation with other parts for purpose of IRC botnet member(s) detection.

### 4.2.2 Phase 2 - Part 2

As stated earlier, Part 2 of Phase 2 will be responsible for detecting the botnet attacks. So, all the outgoing alerts from the internal host(s) with signature names that not included the filtering list and with timestamp longer than or equal *(time_log)* will be collected. These alerts will be filtered in a way to ensure that the collected alerts are achieving the botnet attacking definition. The alerts' attributes in the botnet attack should achieve similar alerts attributes like (timestamp, destination IP, destination port and signature name). That is meaning that the produced alerts are representing attacks from the internal host(s) toward a certain destination (IP and port) that repeated concurrently more than once. In the clustering stage, all the filtered alerts will be clustered according to (source IP and signature name) to extract the hosts and the names of the detected attacks. These hosts will be stored in Report 4 to be correlated with the results of Phase 1 and Phase 2 (Report 1.1, Report 1.2, Report 2, Report 3.1 and Report 3.2).

## 4.3 Spatial-Temporal Correlation and Similarity Engine

The final stage for botnet detection is the spatial-temporal correlation and similarity engine. The stage will deal with six input lines. These lines will be pointed by their reports' numbers (Report 1.1, Report 1.2, Report 2, Report 3.1, Report 3.2 and Report 4). All the reports will be correlated together in a certain way to come up with the final result of botnet detection. The detection results will be divided into three parts based on the obtained information from Phase 1 and number of the bot inside the network, to coherent mode results, non-coherent mode results and one IRC bot results on the standard and non-standard IRC ports.

### 4.3.1 Coherent Mode Correlation Results

The results of coherent mode process flow will be divided to coherent mode on the non-standard IRC ports, coherent mode on the standard IRC ports and the one IRC bot on the non-standard IRC port's results. Figure 2 is showing the correlation process steps for coherent mode and the one IRC bot on the non-standard IRC ports. The whole process of coherent mode correlation results will be clarified as follows:

**1. Correlation of non-standard IRC results**

The results of Report 1.1 that has the initial detected IRC bot(s) of non-standard IRC ports will be correlated with the raw alerts' results of Report 3.1 that has the suspicious IRC responses message on the non-standard IRC ports also. This correlation process will look for similar and continuous C&C responses messages for the initially detected bots in Report 1.1. This objective will be achieved by correlating similar alerts' attributes (source IP, source port, destination IP and destination port) where Report 3.1 has timestamp longer than timestamp on Report1.1. The taken attributes, except timestamp, are representing the bot patterns that can appear for the same IRC bot behavior at different intervals (timestamps). Since every IRC bot will take certain and unique source port on a specific host (source IP) to respond to certain IRC C&C server (destination IP and destination port). The result of this part will produce the detected IRC bot(s) that have connected to a malicious IRC C&C server and still have continued and malicious C&C responses to that server at different time intervals. So, the status field of that bot(s) will be "Coherent mode: IRC bot has illegitimate IRC connection on non-standard IRC port and C&C response(s)". The process of correlation will not stop here, where the produced results from the previous correlation will be further correlated with the results of Report 4. This correlation will look for botnet attacks that could be happened when the bot received a command from the botmaster. So, alerts attribute like (source IP and timestamp) will be considered. This correlation means, if one of the internal hosts (source IP) has been detected performing botnet attack in Report 4, and has an outgoing IRC response message from Report 3.1 at same timestamp; this host will be indexed to be correlated with the results of the previous correlation results. So, if this pattern appeared on the indexed host(s) of the second correlation part; this bot(s) will get status "Coherent mode: IRC bot has illegitimate IRC connection on non-standard IRC port, C&C response(s) and botnet attack". The detected bot(s) in this part will be considered fully utilized IRC bots by the botmaster. Report 1.1 had initial IRC bots and C&C responses. So, the results of Report 1.1 will be compared with the final coherent correlation of non-standard IRC port results; to make sure that all the detected IRC bots in Report 1.1 are not repeated in the last correlation result of non-standard IRC ports. Finally, both of non-standard IRC results and standard IRC results will be grouped as final coherent results before being saved in Report 5, which is representing the final report of IRC botnet detection.

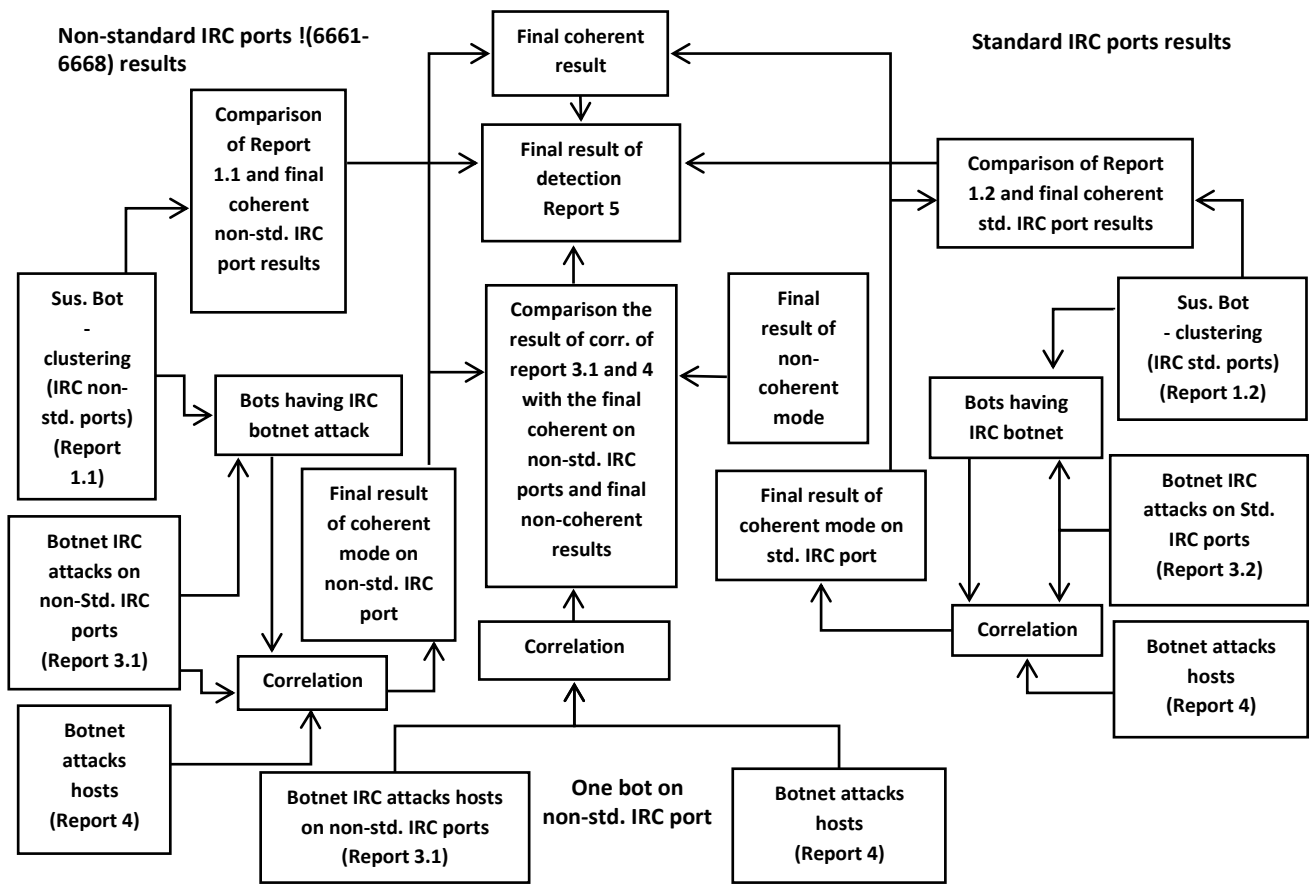

**Fig 2: The coherent mode correlation steps and the one IRC bot on the non-standard IRC ports correlation**

**2. Correlation of standard IRC results**

This kind of correlation will be quite the same as the previous part of correlation but with different reports, where Report 1.2 will be correlated first with Report 3.2, and the result of correlation will be correlated with Report 4. Firstly, The result of Report 1.2 which is representing the initial detected IRC bot on the standard IRC ports will be correlated with the result of Report 3.2, which has the alerts of the IRC responses messages on the same IRC ports. The results of this correlation process will get status message stated that the detected bot(s) in "Coherent mode: IRC bot has illegitimate IRC connection on standard IRC port and C&C response(s)". Secondly, the previous result will be correlated with Report 4 looking for any botnet behavior (botnet attack) for the detected bots. This correlation process is including indexing the host(s) that are performing the botnet attack in Report 4 atthe same timestamp with an IRC response message from Report 3.2. After that, the results of the first correlation will be correlated with the indexed host(s) of each result. The purpose of this correlation is to look for the host of the detected IRC bot. Finally, the detected bot will get this status message "Coherent mode: IRC bot has illegitimate IRC connection on standard IRC port, C&C response(s) and botnet attack".

**3. One bot on non-standard IRC results**

The single botnet member in the monitored network which works on the non-standard IRC port will be detected also. Generally, all IRC bots and even the single bot that have initial IRC activity (NICK) and work on non-standard IRC port should be detected in the coherent mode. However, in the event when the initial activity of the bot is missing, the single bot will have only ordinary C&C responses. The clustering of Report 3.1 which contains all suspicious C&C responses on non-standard IRC port will detect that bot with the status message "IRC bot has illegitimate IRC connection on non-standard IRC port". After that, the correlation of botnet member(s) in Report 3.1 and Report 4 will be achieved by looking for any botnet attack from that bot(s). The detected bots will get the status message "IRC bot has illegitimate IRC connection on non-standard IRC port and botnet attack". The result of this correlation will be compared with the results of the final correlation of coherent mode on non-standard IRC ports, and also with the final results of the non-coherent mode correlation; to make sure that the detected bot has not been detected earlier.

**4.3.2 Non-coherent Mode Correlation**

The results of non-coherent mode process flow will be divided into non-coherent mode on all ports and one bot on the standard IRC ports.

**1. Non-coherent mode on all portsresults**

This part of correlation is quite similar to the previous part (coherent mode); the difference will be on the chosen part from Phase 1. The results of Report 2 will be correlated with its raw alerts to look for similar and continuous C&C responses messages issued from its detected bots. The timestamp value in the raw alerts' report should be longer than timestamp value in Report 2 for the same pattern. So, any duplicated pattern with longer timestamp in the raw alerts' report will state that the detected bot in Report 2 has a continuous C&C responses on the same IRC C&C server. The detected bots will get the status message "Non-coherent mode: IRC bot has illegitimate connection and continues C&C response(s)". The next part of correlation will take the results of the first correlation of non-coherent mode and correlate them with Report 4. Firstly, the raw alerts of Report 2 will be correlated with Report 4 looking for any C&C response in the raw alerts that responsible for any botnet attack in Report 4. The host (source IP) that has the attack at the same timestamp of C&C response will be indexed for the final stage of correlation. The final correlation will be achieved by correlating the first correlation results with the result of the indexed host(s). The bots that are achieving this correlation will get status "Non-coherent mode: IRC bot has illegitimate connection, continues C&C response(s) and botnet attack". The final result of non-coherent correlation will be saved in Report 5. As stated earlier, Report 2 is having a part of IRC botnet detection results of non-coherent mode. So, the results in Report 2 will be compared with the final non-coherent correlation results obtained from the final correlation stage. Finally, the detected IRC bots will be added to Report 5.

**2. One bot on standard IRC results**

Single IRC bot can be detected by correlating results of Report 3.2, which has the alerts of IRC responses messages on the standard IRC ports, along with the results of Report 4 looking for similar alerts' attributes (source IP and timestamp). The chosen attributes will detect all the internal host(s) that has been found having a C&C response at the same timestamp with an outgoing botnet attack. Finally, the results of this correlation will be compared with the results of both non-coherent and coherent mode on the standard IRC results, which stated earlier by checking the similar (source port and timestamp), to ensure that the detected bot had not been detected earlier. As for the status message for detected bots, it will be "IRC bot has illegitimate IRC connection on standard IRC port and botnet attack". As stated earlier, the results of Report 4 in Part 2 of Phase 2 can help a lot in the process of botnet detection regardless of the used botnet communication protocol. So, Report 4 will be included individually in the final results of botnet detection. All the results of Non-coherent mode correlation will be saved in the final detection report (Report 5). Finally, the malicious IRC C&C responses messages would be filtered from the total IRC messages after getting the final results from Report 5. The malicious IRC messages will be filtered by correlating (source IP, source port, destination IP and destination port) between Report 5 and the raw alerts of Report 2 to get only the malicious IRC messages alerts.

## 5. Evaluation

This section will prove the efficiency and the accuracy of the proposed model to detect the IRC botnet member(s) and their behaviors. In fact, there will be two conducted case studies; first one is on Virtual network with multiple botnet infection scenarios. The second one will be on DARPA 2000 - Windows NT Attack Data Set that contains normal IRC along with Windows NT attack. Based on the obtained results, the comparison will be between the proposed model and some of the well-known approaches in the field of IRC botnet detection, including BotHunter[23], BotSniffer[7] and Rishi [14]. The benchmark of the comparison will be in terms of the botnet characteristics for the chosen approaches.

### 5.1 Building Experiment

As stated earlier, the proposedmodel depends on Snort sensor to generate alerts on the suspicious activities. So, Snort IDS should be installed on the system that is scheduled to be the network monitoring system. The proposed model will take the raw alerts and analyze them to produce the detected botnet member(s). Nmap, the network scanning tool, also will be used to collect all the hosts inside the monitoring network. The chosen system to achieve the experiment is Unix-based system (Ubuntu Linux) installed with all needed tools for the experiment. These tools will include Apache as a web server, PHP and MySQL for implementing the proposed model. In the experiment, the monitored network will be on a virtual network which is constructed inside VMware Workstation, and the Snort machine will be one of them. Figure 3 is depicting clearly the design of the testing environment.

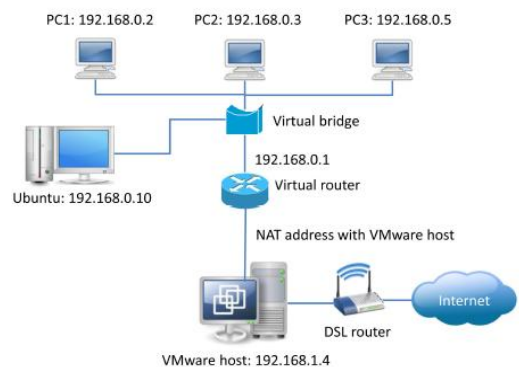

**Fig3: VMware testing environment**

### 5.1.1 IRC Bots

There are two different types of IRC bots will be exploited in the experiment.One of them is considered common and well-known type of IRC bot which is Rxbot and the other one is a new one, which is CRIME SCENCE bot. CRIME SCENE Bot has been developed recently byJarad G.[24]. Of the most important features of this bot is that it has been written in three different types of programming languages like (C++, C# and Python). CRIME SCENE Bot has a lot of other useful features to the botmaster like Email notification and FTP connectivity to upload the leaked information.

### 5.2 Evaluation Case Studies

There will be two case studies to be conducted to evaluate the efficiency and the accuracy of the proposed model. These case studies will represent multiple botnet scenarios for the IRC bots in different situations and behaviors.

### 5.2.1 Case Study 1: Virtual Network Experiment

This case study will be divided into multiple scenarios to represent different situations of existing botnet inside the monitored network with different intervals of monitoring time. The taken procedures of these scenarios will be clarified as follows:

**1. Botnet Scenario 1: Detecting Multiple Kinds of Bots in Initial IRC Activity**

The first objective of this scenario is to show the efficiency of the proposed model to detect IRC bots that have initial activity (NICK messages).The second objective of this scenario is to prove that the proposed model can detect IRC botnet member(s) that working on different IRC destination ports and IPs. The internal hosts of the network have been infected manually with five bots (two CRIME SCENE bots in PC2 and PC3 and three RxBot bots in PC1, PC2 and PC3). RxBot type will connect to local C&C server on standard IRC port 6667 and CRIME SCENE bots will connect to remote C&C server on non-standard IRC port 7000.

**2. Botnet Scenario 2: Detecting Multiple Kinds of Bots in Middle of IRC Activity**

The first objective of this scenario is to show the efficiency of the proposed method to detect botnet members that do not have any initial activity (NICK) and do not have any kind of detectable network attacks like DDoS attack. The second objective is to evaluate the detection efficiency after updating the Snort signature rules with rules from Bleeding Snort web site [22]. The third objective is to evaluate the efficiency of the proposed model to filter out the normal IRC messages toward single IRC server (same destination IP and destination port) without any false positive results for botnet detection. To achieve this situation in the experiment, four bots in different situations will be used (two CRIME SCENE bots with initial activities and two RxBot bots without initial activities).

**3. Botnet Scenario 3: Detecting Single Bot Members at Different Situations**

The objective of this IRC botnet scenario is to show the efficiency of the proposed model to detect single botnet members from different botnets at different situations and behaviors inside the monitoring network. To achieve this situation in the experiment, three bots have been used, two bots from CRIME SCENE (one of them has the initial IRC bot activity and other has not) with one RxBot bot without initial IRC activity.

### 5.2.2 Case Study 2: Windows NT Attack DARPA 2000 Network Traffic Data Set

DARPA (Defense Advanced Research Project Agency) 2000 is an intrusion-detection evaluation data set contains multiple network attacking scenarios. This data set is mainly designed to evaluate the detection probability and false detection probability for the network security system under test, especially in the intrusion-detection research field [25]. In this paper, Windows NT Attack Data Set from DARPA 2000 will be used to evaluate the proposed model. The main objective will be to show that the proposed model can pass through the normal IRC chatting, which is happening on standard IRC port 6667 without any false positive result of IRC botnet inside the network. This data set contains two flows from two networks; one for the Inside and one for Outside network.

### 5.3 Results and Discussion

The obtained results will be discussed by reviewing the total number of alerts that obtained from each experiment with the percentages of the IRC messages alerts (normal and malicious). The proposed model will be evaluated also regarding to how the proposed model could achieve the proposed objectives and state the current status of the detected bots besides of detecting their behaviors. The total number of alerts could be varied in thestated experiments from the others' experiments, since the default Snort rules set has been updated with the enhanced Snort rules, and with rules set from Bleeding Snort rules. The proposed model will count the total alerts, IRC message alerts and the percentages of the normal and malicious IRC messages.

### 5.3.1 Case Study 1: Virtual Network Experiment Results

As stated earlier, the first case study is including three different botnet scenarios, where each one comes with different botnet situations. The results and discussion of each scenario will be addressed as follows:

**1. Botnet Scenario 1: Detecting Multiple Kinds of Bots in Initial IRC Activity Results and Discussion**

Table 1 shows the obtained details from the result of first botnet scenario conducted in the experiment. As depicted in table 1, the proposed model could extract the following results after detecting the IRC bots: The total IRC messages are only occupying a percentage of 3% (324 alerts) of the total alerts (10,259), and 56% (182 alerts) of them are malicious and the rest normal messages are 44% (142 alerts). Since there were 5 IRC bots with three legitimate IRC clients inside the monitored network.

**Table 1:Results of IRC Botnet Scenario 1**

| T-A | T-I | P-I-A | N-I | M-I | P-N-I | P-M-I | I-B | D-I-B |
|---|---|---|---|---|---|---|---|---|
| 10,259 | 324 | 3% | 142 | 182 | 44% | 56% | 5 | 5 |

Where: **T-A**: Total No. Alerts, **T-I**: Total No. of IRC messages, **P-I-A**: Percentage of IRC messages to the total No. of Alerts, **N-I**: No. of Normal IRC messages, **M-I**: No. of Malicious IRC messages, **P-N-I**: Percentage of Normal IRC messages to total No. of IRC messages, **P-M-I**: Percentage of Malicious IRC messages to total No. of IRC messages, **I-B**: No. of IRC Bots inside the network and **D-I-B**: No. of Detected IRC Bots.

By analyzing the IRC messages' alerts and botnet attacks, the proposed model could prove the following objectives:

1. The proposed model was accurate to detect all the infected IRC bots (the five bots). The accuracy of detection was 100%. As for the accuracy of stating the current status for the detected bots, it was 80%. Since there was only one bot out of the five bots, has not gotten its current status.

*International Journal of Computer Applications (0975 – 8887) Volume 66– No.15, March 2013*2. The proposed model could detect IRC botnet members that work in different IRC ports with different destination IP addresses for IRC C&C server.

As for the filtered IRC malicious and normal alerts, the proposed model could filter them accurately after detecting the IRC bots as shown in table 1.

**2. Botnet Scenario 2: Detecting Multiple Kinds of Bots in Middle of IRC Activity Results and Discussion**

Table 2 shows the details of the results obtained from conducting the second botnet scenario. As shown in the table, the total number of alerts is 2,073, and it is less than the total number of alerts in the previous botnet scenario. This is because the time interval of the second scenario experiment was less than the first one, there was no DDoS attack in this scenario and also, the number of IRC bots is less by one bot from the previous scenario. There were 8% (163 alerts) of the total alerts related to IRC messages, 41% (67 alerts) of them were malicious C&C responses messages. Since there are four bots in different situations inside the monitored network. As for the normal IRC chatting, there were three normal IRC clients, and they had 59% (96 alerts) of the total IRC messages and all of them were directed to single IRC chatting server.

**Table 2:Results of IRC Botnet Scenario 2**

| T-A | T-I | P-I-A | N-I | M-I | P-N-I | P-M-I | I-B | D-I-B |
|---|---|---|---|---|---|---|---|---|
| 2,073 | 163 | 8% | 96 | 67 | 59% | 41% | 4 | 4 |

So, it has been shown from the above that the conducting of this scenario proved the following objectives:

1. The proposed model was 100% accurate to detect the infected IRC bots in different situations (coherent mode or non-coherent mode).

2. The added enhanced rules set could tag the destination port for the IRC C&C server of the botmaster. However, the tagging process may not work during the whole experiment interval or may not working at all, due to functional conflict between the rules and the Bleeding Snort rules for IRC message monitoring.

3. The proposed model was 100% accurate to filter out the C&C responses messages from normal IRC messages even when the normal messages have single IRC server for destination IP and destination port just like the server of the IRC bot.

**3. Botnet Scenario 3: Detecting Single Bot Members at Different Situations Results and Discussion**

Table 3shows the details of the results obtained from conducting the experiment of botnet scenario 3. As depicted in the table, the total alerts obtained from the result were 60,989 alerts and most of them were belonged to ICMP DDoS attack. As for IRC traffic, there were only 149 alerts for the IRC the messages. As stated earlier in the experiment of botnet scenario 3, there were three bots in different situations inside the monitored network, and they had 34% (50 alerts) of the total IRC messages as IRC responses messages. As for the normal IRC chatting messages, they took 66% (99 alerts) from the total IRC messages.

**Table 3 :Results of IRC Botnet Scenario 3**

| T-A | T-I | P-I-A | N-I | M-I | P-N-I | P-M-I | I-B | D-I-B |
|---|---|---|---|---|---|---|---|---|
| 60,989 | 149 | 0.2% | 99 | 40 | 66% | 34% | 3 | 3 |

So, it has been shown from the above that the conducting of this scenario proved the desired objective of detecting 100% of the infected singles IRC bot(s). Even when they were in different situations with different activities (botnet attacks).

**5.3.2Case Study 2: Windows NT Attack DARPA 2000 Data set Results and Discussion**

Table 4 shows the results information of the conducted experiment of case study 2 for Windows NT Attack Inside tcpdump file. The total number of alerts was 930 alerts for the whole interval time of monitoring, which was about 5.5 hours. There were 35 IRC messages (4% of the total alerts) during the experiment, and all of them were normal IRC messages. As for botnet behavior detection, the Inside network data set testing archived one host performing two false positive botnet behaviors out of 40 hosts. These false positive results happened, since the detected activities are quite similar to the botnet attack's entropy. However, the false positive alerts could be filtered easily by adding the detected alerts to the filtering alerts of botnet behaviors part in the proposed model to pass these alerts in the next time analysis.

**Table 4:Results of case study 2 for the Inside tcpdump file**

| T-A | T-I | P-I-A | N-I | M-I | P-N-I | P-M-I | I-B | D-I-B |
|---|---|---|---|---|---|---|---|---|
| 930 | 35 | 4% | 35 | 0 | 100% | 0% | 0 | 0 |

Table 5 shows the details of the results obtained from conducting the experiment of analyzing Windows NT Attack Outside tcpdump file. As depicted in the table, the total number of alerts was 481 and only 11% (53 IRC messages) of it was for IRC messages. However, this data set also does not contain any kind of botnet but it was tested to evaluate the proposed model ability to pass the normal IRC messages. The Outside network data set testing did not achieve any false positive results regarding to botnet behavior detection.

**Table 5:Results of case study 2 for the Outside tcpdump file**

| T-A | T-I | P-I-A | N-I | M-I | P-N-I | P-M-I | I-B | D-I-B |
|---|---|---|---|---|---|---|---|---|
| 481 | 53 | 11% | 53 | 0 | 100% | 0% | 0 | 0 |

The objective from the previous two experiments has been achieved and showed that the proposed model was accurate 100% to pass the legitimate IRC messages without any false positive results in IRC botnet detection. The conducting of the experiment also shows some of the false positive results regarding to botnet behaviors (attack) detection.

## 6. Comparison with other Botnet Detection Approaches

It is hard to conduct a fair comparison between the existing approaches for botnet detection and the proposed model. The reason is due to many factors, including;

1. Every exiting approach has been evaluated into different network environment [26].
2. Different binary bots have been used in the experiments [26].
3. It is not easy to get and execute the binary code for each approach.

So, the comparison will be in terms of the botnet characteristics for the chosen approaches.

### 6.1 Comparison by Botnet Characteristics

The comparison will based on the botnet characteristics. These characteristics and the botnet approaches evaluations have been introduced byStinson and Mitchell in [17]. Table 6 shows some of the selected botnet characteristics that will be used in the comparison.

**Table 6: Comparison of the proposed model with the other approaches based on botnet**

| Characteristic | Description |
|---|---|
| Basis | Type of the detection method whether host based or network-based. |
| IRC | Depending on specific IRC port number or specific model of communications patterns. |
| Flow characteristics | Depending on certain flow characteristics to correlate C&C communication or/and attacks. |
| Time | Using time window to correlate the network events. |
| Network-based detection | Depending on network-based detection attacksuch as DDoS attack. |
| Syntax | Depending detecting special command, nicknameor protocol syntax. |

**Table 7. Comparison of the proposed model with the other approaches based on botnetCharacteristics**

| Approach | Basis | IRC | Flow-chars. | Time | Net-based | Syntax |
|---|---|---|---|---|---|---|
| Bothunter | Net- | Yes | No | Yes | Yes | Yes |
| Botsniffer | Net- | Yes | No | Yes | Yes | Yes |
| Rishi | Net- | Yes | No | No | No | Yes |
| The proposed model | Net- | Yes | Pps | No | Yes | Yes |

Table 7 shows the comparison of the proposed model with the comparative approaches. This comparison stated that the proposed model is not depending on a specific time window; instead of that IRC C&C responses will be correlated into two ways. The first kind of correlation will be for the C&C responses on non-standard IRC ports, where general clustering approach will be used for clustering the C&C responses messages. The second kind will be for the standard IRC ports, where the packet per second (pps) characteristic will be used for filtering the C&C responses of the IRC bot. Pps also has been used in the correlation of the outgoing C&C response that is related to an outgoing botnet attack. This correlation method has been used to detect the single bot on the standard IRC ports and to correlate the C&C responses behavior with the botnet attacks. The comparison shows also that Rishi has the minimum dependency over the all approaches, but the fact is Rishi can be evaded easier than the other approaches since it is depending on certain predefined templates of suspicious nicknames [14][17]. Bots nicknames could be varied from bot to bot, so Rishi may not have their nickname's pattern on its templates, rather than the false positive results that could be produced from using such approach [14].

## 7. Conclusion

In this paper, a multi-phase model for detecting the IRC botnet and botnet behavior has been proposed. The detection method depends on the outgoing IRC C&C responses messages between the bots and their botmaster. After conducting several different experiments to evaluate the proposed model efficiency, the conclusion is that the proposed model could detect even new and single IRC bots like (CRIME SCENE IRC bot) which has a random responding time. The malicious IRC C&C responses messages can be filtered out from the normal ones as a result of detecting their sources (the IRC bots). Botnet behaviors like DDoS attacks could be verified also by the proposed model. So these behaviors could indicate that the attacking host(s) is a botnet member(s) regardless of the used protocol for C&C instructions. The proposed model still not the complete and perfect solution for IRC botnet and botnet behavior detection problem. There are some situations make the model fail to detect the bots totally or partly. The bots that used encrypted IRC channel will not be detected, but the proposed model still able to detect the IRC botnet attacks as long as Snort can detect them. The proposed model could be evaded if the attacker knows the internal structure of the proposed model. The attacker can run the bots on the defined range of the standard IRC ports and make the bots responding time in a random way to break the pps correlation method. In this situation, the whitelist strategy can help to filter the IRC traffic of a certain application when the network administrator chooses to modify the default standard IRC ports on the proposed model. Finally, the proposed model is not designated to prevent IRC botnet attacks from happening, as is the case in IDS. But still interesting for the network administrator to see the IRC traffic analysis for botnet detection, to take an action based on the results' analysis immediately.

## 8. REFERENCES

[1] Zhuge, J., Holz, T., Han, X., Guo, J. and Zou, W. 2007. Characterizing the IRC-based botnet phenomenon, Technical report, Peking University , University of Mannheim.

[2] Westervelt, R. 2011.Waledac Botnet showing resurgence with thousands of stolen email credentials. URL: http://searchsecurity.techtarget.com/news/1527003/Waledac-botnet-showing-resurgence-with-thousands-of-stolen-email-credentials,.

[3] Neil, D., Stoppelman and Michael. 2007. The anatomy of clickbot.A, Proceedings of the first conference on First Workshop, Hot Topics in Understanding Botnets, USENIX Association, Berkeley, CA, USA, pp. 11–11.

[4] Baecher, P., Koetter, M., T. Holz, M. D. and Freiling, F. 2006. The nepenthes platform: An efficient approach to collect malware, in K. C. Zamboni, Diego (ed.), Recent Advances in Intrusion Detection, Vol. 4219 of Lecture Notes in Computer Science, Springer Berlin, Heidelberg, pp. 165–184.

[5] Ioannidis, J. and Bellovin, S. M. 2002. Implementing pushback: Router-based defense against DDoS attacks, Proceedings of Network and Distributed System Security Symposium. Network and Distributed System Security Symposium: NDSS '02 (Reston, Va.: Internet Society).

[6] Zeidanloo, H. R. and Manaf, A. B. A. 2010. Botnet detection by monitoring similar communication patterns, Journal of Computer Science 7(3): 36–45. URL: http://arxiv.org/abs/1004.1232,.

[7] Gu, G., Junjie, Z. and Wenke, L. 2008. Botsniffer : Detecting botnet command and control channels in network traffic, Technology 53(1): 1–13.

[8] Grizzard, J. B., Sharma, V., Nunnery, C., Kang, B. B. and Dagon, D. 2007. Peer-to-peer botnets: overview and case study, Proceedings of the first conference on First Workshop on Hot Topics in Understanding Botnets, USENIX Association, Berkeley, CA, USA, pp. 1–1. URL: http://dl.acm.org/citation.cfm?id=1323128.1323129,.

[9] Micro, T. 2006. Taxonomy of botnet threats, Malware White Papers pp. 1–15. URL: http://www.webbuyersguide.com/resource/resourceDetails.aspx?id=8021,.

[10] Zeidanloo, H.R., Shooshtari, M., Amoli, P., Safari, M. and Zamani, M. 2010. A taxonomy of botnet detection techniques, Computer Science and Information Technology (ICCSIT), 3rd IEEE International Conference on, IEEE Computer Science and Information Technology, Chengdu, China, pp. 158 – 162.

[11] Honeynet Project 2006. Know your enemy: Honeynets, URL: http://www.honeynet.org/papers/honeynet,.

[12] Baecher, P., Koetter, M., T. Holz, M. D. and Freiling, F. 2006. The nepenthes platform: An efficient approach to collect malware, in K. C. Zamboni, Diego (ed.), Recent Advances in Intrusion Detection, Vol. 4219 of Lecture Notes in Computer Science, Springer Berlin, Heidelberg, pp. 165–184.

[13] Zhichun, L., Anup, G. and Yan, C. 2008. Honeynet-based botnet scan traffic analysis, in W. C. D. D. Lee, Wenke (ed.), Botnet Detection, Vol. 36 of Advances in Information Security, Springer US, pp. 25–44.

[14] Goebel, J. and Holz, T. 2007. Rishi: identify bot contaminated hosts by IRC nickname evaluation, Proceedings of the first conference on First Workshop on Hot Topics in Understanding Botnets, USENIX Association, Berkeley, CA, USA, pp. 8–8. URL: http://dl.acm.org/citation.cfm?id=1323128.1323136,.

[15] Liu, H., Sun, Y., Valgenti, V. C. and Kim, M. S. 2011. Trustguard: A flow-level reputation-based DDoS defense system, Consumer Communications and Networking Conference (CCNC), Washington State University, IEEE, Pullman, Washington 99164-2752, U.S.A.

[16] Binkley, J. R. and Singh, S. 2006. An algorithm for anomaly-based botnet detection, Vol. 06, USENIX Association, pp. 43–48.

[17] Stinson, E. and Mitchell, J. 2007. Characterizing bots remote control behavior, in B. M. Hammerli and S. Robin (eds), Detection of Intrusions and Malware, and Vulnerability Assessment, Vol. 4579 of Lecture Notes in Computer Science, Springer Berlin, Heidelberg, pp. 89–108.

[18] Gu, G., Yegneswaran, V., Porras, P., Stoll, J. and Lee, W. 2009. Active botnet probing to identify obscure command and control channels, Proceedings of Annual Computer Security Applications Conference (ACSAC'09).

[19] Dingbang, X. and Peng, N. 2008. Correlation analysis of intrusion alerts, Intrusion Detection Systems, Vol. 38 of Advances in Information Security, Springer US, pp. 65–92.

[20] Bailey, M., Cooke, E., Jahanian, F., Xu, Y. and Karir, M. 2009. A survey of botnet technology and defenses, 2009 Cybersecurity Applications Technology Conference for Homeland Security (01): 299–304. URL: http://ieeexplore.ieee.org/lpdocs/epic03/wrapper.htm?arnumber=4804459,.

[21] Bianco, D. 2006. Detecting common botnets with snort, URL: http://blog.vorant.com/2006/03/detecting-common-botnets-with-snort.html,.

[22] Bleeding Snort 2006. Bleeding snort website, URL: http://www.bleedingsnort.com,.

[23] Gu, G., Porras, P., Yegneswaran, V., Fong, M. and Lee, W. 2007. Bothunter: Detecting malware infection through ids-driven dialog correlation, Proceedings of the 16th USENIX Security Symposium (Security'07).

[24] Grant, J. 2012. IRC bot source codes (written in C++, C# & Python), URL: http://darksidegeeks.com/irc-bot-source-codes-c-c-python,.

[25] MITL Lab. 2000 DARPA intrusion detection scenario-specific datasets, URL: http://www.ll.mit.edu/mission/communications/ist/corpora/ideval/data/2000data.html,.

[26] Lu, W., Rammidi, G. and Ghorbani, A. A. 2011. Clustering botnet communication traffic based on n-gram feature selection, Computer Communications 34(3): 502 – 514. URL: http://www.sciencedirect.com/science/article/pii/S0140366410001751,.